\documentstyle[12pt]{article}

 \newcommand{\be}{\begin{equation}}
 \newcommand{\ee}{\end{equation}}
 \newcommand{\ba}{\begin{eqnarray}}
 \newcommand{\ea}{\end{eqnarray}}
 \newcommand{\spa}{\hspace{0.5cm}}
 \newcommand{\del}{\partial}

\newcommand{\tende}{\stackrel{|\vec x-\vec y|\rightarrow \infty}
{\longrightarrow}}

\newcommand{\tendeo}{\stackrel{\rho, \delta \rightarrow 0}
{\longrightarrow}}

\def\overbar{\bar}

\def\A{\tilde A}

\def\B{\tilde B}

\newcommand{\expo}{\exp \lef \{ - \int d^3z }

\newcommand{\lef}{\left}

\newcommand{\ri}{\right}

\newcommand{\cl}{{\cal L}}

\newcommand{\fr}{\frac}

\begin{document}

\begin{titlepage}

\topmargin -15mm

\vskip 10mm

\centerline{ \LARGE\bf A New Quantum Vortex Operator }
\vskip 3pt
\centerline{\LARGE\bf and Its Correlation Functions $^*$}

    \vskip 2.0cm

    \centerline{\sc E.C.Marino }

     \vskip 0.6cm

    \centerline{\it Instituto de F\'\i sica }
    \centerline{\it Universidade Federal do Rio de Janeiro }
    \centerline{\it Cx.P. 68528}
    \centerline{\it Rio de Janeiro RJ 21945-970}
    \centerline{\it Brasil}

\vskip 1.0cm

\vskip 1.0cm

\begin{abstract}

A new local and gauge invariant quantum vortex operator is constructed
in three-dimensional gauge field theories. The correlation functions of this
operator
are evaluated exactly in pure Maxwell theory and by means of a loop
expansion in the Abelian Higgs model. In the broken symmetry phase of the
latter an explicit expression for the mass of the quantum vortices
is obtained from the long distance exponential decay of the two-point
function.

\end{abstract}

\vskip 6cm
$^*$ Work supported in part by CNPq-Brazilian National Research Council.
     E-Mail address marino@if.ufrj.br

\end{titlepage}

\hoffset= -10mm

\leftmargin 23mm

\topmargin -8mm
\hsize 153mm

\baselineskip 7mm
\setcounter{page}{2}

\section{Introduction}

\spa In several domains of physics a thorough quantum description of
topological excitations is frequently needed. This is the case, for
instance, of the solitons of nonlinear optics, vortices in superfluids
and superconductors, magnetic monopoles and cosmic strings, to cite some
examples. In these cases it is highly desirable, not only from the point
of view of basic principles and aesthetics, but also from a practical and
calculational perspective, to have a theory dealing directly with the
quantum creation operators of the relevant topological excitations. A
common feature of such kind of excitations is that they cannot in general
be expressed polinomially in terms of the basic lagrangian variables and
therefore a fundamental question is how to construct the corresponding
creation operators.

\spa In the present work we are going to concentrate on vortices of
2+1 dimensional gauge field theories. A quantum vortex operator was
constructed in continuum space-time \cite{vor} some years ago and its
correlation functions were evaluated within some approximation scheme
\cite{vorcorr}. Other similar constructions appeared previously on the
lattice \cite{rede} and subsequently on the continuum \cite{cont}
 A general analysis of these operators in several
gauge theories in 2+1D was performed more recently \cite{an}, paying
special attention to the case of Chern-Simons theories. In this paper,
we will obtain a new quantum
vortex operator which presents many advantages with respect to the former
one \cite{vor} and even corrects some  defects associated with the previous
formulation \cite{vor,evora}. The new operator is,
for instance, explicitly surface
invariant and gauge invariant, even in the presence of a coupling to a
Higgs field. There are no cutoffs appearing in the definition of the operator
which should be removed at the end of any calculation. In the nontrivial
case of the Abelian Higgs model, locality of the correlation function is
attained already in the lowest order of a loop expansion,
through an exact cancellation of the nonlocal terms. Throughout the paper,
we will comment more on each of the several advantages of the
present formulation.

\spa In Sect. 2 we start by obtaining an explicitly surface invariant
quantum vortex operator in the pure Maxwell theory. This operator is then
generalized for the case of the Abelian Higgs Model where an  explicitly
gauge invariant form is obtained. The
relevant commutation rules are evaluated at the end of the section, both in
the Maxwell and Abelian Higgs theories.

\spa In Sect. 3 we evaluate the correlation functions of the new vortex
operator, firstly in Maxwell theory, where an exact result can be obtained
but there are no genuine vortex excitations and then in the Abelian Higgs
model. In this case a loop expansion is made in order to obtain the long
distance behavior of the correlation function. Contrary to the previous
formulation \cite{vorcorr}, a mass expansion is
no longer needed for the obtainment of
the correlation function. In the broken symmetry phase there are true
quantum vortex excitations whose mass is explicitly obtained from the
long distance behavior of the two-point correlation function.

\section{The Vortex Operator and Its Commutation Rules }

\setcounter{equation}{0}

\subsection{The Maxwell Theory}

\spa In previous works, a quantum vortex operator was obtained in 2+1D
\cite{vor,an} for theories involving an abelian gauge field.
In terms of the electric field $E^i=F^{i0}$, it can be expressed
as
\be
\mu (x) = \exp \lef\{ - ib \int_{T_x(C)} d^2 \vec{\xi} F^{i0}(\vec\xi,x^0)
   \del_i arg(\vec\xi - \vec x) \ri \}
\label{2.1}
\ee
where b is an arbitrary real parameter with
  dimension $mass^{-1/2}$ and $T_x(C)$ is the surface
represented in Fig.1.

 \spa Let us introduce now the new
external field
\be
\tilde A_{\mu\nu}(z;x) = b \int_{T_x(C)} d^2\xi_\nu \del_\mu
arg(\xi -x) \delta^3(z-\xi) - (\mu \leftrightarrow \nu)
\label{2.2}
\ee
where $ d^2\xi_\nu$ is the surface element of the surface $T_x(C)$. We
can express the operator$\mu$ in terms of this and the
field intensity tensor as
\be
\mu(x) = \exp \lef\{ - i \int d^3z \tilde A_{\mu\nu} F^{\mu\nu} \ri\}
\label{2.3}
\ee
In order to get local
correlation functions for
$\mu$, one has to introduce a c-number renormalization factor which
corresponds to the self-coupling of the external field $\tilde A_{\mu\nu}$.
Let us start by considering the pure Maxwell theory. In this case,
\be
<\mu(x)\mu^\dagger(y)>=Z^{-1} \int DA_\mu \expo \lef [ \fr{1}{4}
(F_{\mu\nu}+ \A_{\mu\nu})(F^{\mu\nu}+\A^{\mu\nu}) \ri ]\ri\}
\label{2.4}
\ee
Observe that since $\A_{\mu\nu}$ is not in the form $\del_{[\mu}\A_{\nu]}$
we cannot eliminate the external field from (\ref{2.4}) by a shift in the
functional integration variable as could be done in \cite{vor,vorcorr}.
The present formulation, therefore corrects this defect of the previous one.

\spa In order to show that the above expression indeed does not depend on the
specific choice of the surface $T_x(C)$, let us choose another arbitrary
surface $\tilde T_x(C)$, also bonded to the curve $C_x$ according to Fig.2.
Let us then perform the following change in the functional
integration variable
in (\ref{2.4})
\be
A_\mu \rightarrow A_\mu + \Lambda_\mu
\ee
with
\be
\Lambda_\mu= -b \Theta (V(\tilde T_x)) \del_\mu arg(\vec z-\vec x)
\label{2.5}
\ee
In this expression, $ \Theta (V(\tilde T_x))$ is the three-dimensional
Heaviside function with support inside the volume $V(\tilde T_x)$ bounded
by $T_x \cup \tilde T_x$.
It is easy to see that because the singularity of the $arg$ function is
outside this volume, we have, under the above change of variable
\be
F_{\mu\nu} \rightarrow F_{\mu\nu} -b \del_\mu   \Theta (V(\tilde T_x)\del_\nu
arg(\vec z-\vec x) - (\mu \leftrightarrow \nu)
\ee
or
$$
F_{\mu\nu} \rightarrow F_{\mu\nu} -b \oint_{\tilde T_x -T_x}
d^2\xi_\mu \del_\nu arg(\vec z-\vec x) \delta^3(z-\xi) -
(\mu \leftrightarrow \nu)
$$
\be
F_{\mu\nu} \rightarrow F_{\mu\nu} + \A_{\mu\nu}(\tilde T_x) - \A_{\mu\nu}(T_x)
\label{2.6}
\ee
{}From (\ref{2.4}) and (\ref{2.6}) it is clear that the correlation function
$<\mu\mu^\dagger>$ is surface invariant.
As was shown in \cite{vorcorr}, removal
of the cutoffs $\rho$ and $\delta$ immediately leads to local correlation
functions.

\spa As we will demonstrate now, it is possible to
obtain a simpler form for the
vortex operator $\mu$, in which the presence of
no cutoff is needed. In order to achieve this, let us start by performing
in (\ref{2.4}) the following change of functional integration variable
\be
A_\mu \rightarrow A_\mu + \A_\mu(z;x,y)
\label{2.13}
\ee
where
\be
\A_\mu (z;T_x) = b \int_{T_x} d^2\xi_\mu arg(\xi -x) \delta^3(\xi -x)
\label{2.11}
\ee

The correlation function becomes
\be
<\mu(x)\mu^\dagger(y)>=Z^{-1} \int DA_\mu \expo \lef [ \fr{1}{4}
(F_{\mu\nu}+ \overbar A_{\mu\nu})(F^{\mu\nu}+\overbar A^{\mu\nu}) + \cl_{GF}
 + V(\phi)\ri ]\ri\}
\label{2.14}
\ee
where
$$
\overbar A_{\mu\nu} = \A_{\mu\nu} - \del_\mu \A_\nu -\del_\nu\A_\mu
$$
\be
\overbar A_{\mu\nu} =
 b \int_{T_x(C)} d^2\xi_\nu \del_\mu
arg(\xi -x) \delta^3(z-\xi)
- \int_{T_x(C)} d^2\xi_\nu arg(\xi -z) \del^{(\xi)}_\mu \delta^3(z-\xi)
 - (\mu \leftrightarrow \nu)
\label{2.15}
\ee
The vortex operator becomes
\be
\mu(x) = \exp \lef\{ - i \int d^3z \overbar A_{\mu\nu} F^{\mu\nu} \ri\}
\label{2.16}
\ee
and we may, therefore, write now
\be
\mu (x) = \exp \lef\{ - ib \int_{T_x(C)} d^2 \vec{\xi}\del_i
\lef [ F^{i0}(\vec\xi,x^0)
    arg(\vec\xi - \vec x)\ri ]
  \ri \}
\label{2.17}
\ee
Observe that this reduces to (\ref{2.1})
because in the absence of matter $\del_i F^{i0}=0$.

\spa Let us take the general expression for the vortex operator, given by
(\ref{2.17}). Observing that we have the surface integral of a total
derivative, we may use Stokes' theorem to write
\be
\mu (x) = \exp \lef\{ - ib \oint_{C_x} d \xi^i  \epsilon^{ij}
 F^{j0}(\vec\xi,x^0)
    arg(\vec\xi - \vec x)
  \ri \}
\label{2.18}
\ee
where $C_x$ is the curve defined in Fig. 1. Taking the limit in which
the cutoffs $\rho$ and $\delta$ go to zero, we can express the vortex
operator in a simple form, in terms of a line integral
\be
\mu (x) = \exp \lef\{ - i 2\pi b \int_{\vec x,L}^\infty d \xi^i
\epsilon^{ij}
 F^{j0}(\vec\xi,x^0)
  \ri \}
\label{2.19}
\ee
This can also be put in a covariant form, namely,
\be
\mu (x) = \exp \lef\{ - i \pi b \int_{\vec x,L}^\infty d \xi^\mu
\epsilon_{\mu\alpha\beta}
 F^{\alpha\beta}(\xi,x^0)
  \ri \}
\label{2.20}
\ee
This form of the vortex operator is a natural extension for 2+1D of
the corresponding soliton operator first introduced in 1+1D by
Mandelstam \cite{man,evora}, namely
\be
\mu_{1+1} = \exp \lef \{ - i b \int_x^\infty d \xi^\mu
\epsilon_{\mu\nu} \del^\nu \phi \ri \}
\label{sol}
\ee
where $\phi$ is a scalar field, the Sine-Gordon field, for instance.

\spa Finally, it is extremely convenient, again, to express the vortex
operator in terms of an external field
\be
\mu (x) = \exp \lef\{ \fr{1}{2} \int d^3z  \tilde B_{\alpha\beta}
 F^{\alpha\beta}(z)
  \ri \}
\label{2.21}
\ee
where the external field $\tilde B_{\mu\nu}$ is given by
\be
\tilde B_{\alpha\beta} = i 2\pi b \int_{\vec x,L}^\infty d \xi_\mu
\epsilon^{\mu\alpha\beta} \delta^3(z-\xi)
\label{2.22}
\ee
It is obvious that we can perform the same transformations we did in
order to arrive at the above expression, also to the quadractic self-
coupling $\bar A_{\mu\nu}$ term in the correlation function. We can
therefore substitute $\bar A_{\mu\nu}$ for $\tilde B_{\mu\nu}$ in
(\ref{2.14}). Observe that also here we cannot eliminate $\tilde
B_{\mu\nu}$ from (\ref{2.14}) by a shift in the functional integration
variable.

\subsection{ The Abelian Higgs Model}

\spa One has to consider now what happens when the gauge field is coupled
to a charged field. Let us take the Abelian Higgs model as a paradigm.
This is defined by
\be
\cl_{AH} = -\fr{1}{4} F^{\mu\nu}F_{\mu\nu} + |D_\mu \phi|^2 - V(\phi)
\label{2.7}
\ee
 As usual, $D_\mu =\del_\mu+ ieA_\mu$, where $e$ is
the gauge coupling constant.

Before considering the construction of the vortex operator in the Abelian
Higgs Model, let us introduce the following representation of the Higgs field
which is going to be very convenient
\be
\phi = \fr{1}{\sqrt 2} \rho  \ {\rm e}^{i\theta}
\label{exp}
\ee
In terms of $\rho$ and $\theta$ we can rewrite the lagrangian as
\be
\cl_{AH} =  -\fr{1}{4} F^{\mu\nu}F_{\mu\nu} + \fr{1}{2} e^2 \rho^2 (A_\mu +
\fr{1}{e} \del_\mu \theta)(A^\mu +\fr{1}{e} \del ^\mu \theta) +
 \fr{1}{2} \del_\mu \rho \del^\mu \rho - V(\rho)
\label{lro}
\ee
This is clearly invariant under gauge transformations
$$
A_\mu \rightarrow A_\mu + \del_\mu \Lambda
$$
\be
\theta \rightarrow \theta - e \ \Lambda
\label{gat}
\ee
Introducing a new vector field
\be
W_\mu = A_\mu + \fr{1}{e} \del_\mu \theta
\label{w}
\ee
and using the field equation for $\theta$
\be
\Box \theta = -e \del_\mu A^\mu
\label{teq}
\ee
which allows us to write
\be
\del_\mu W^\mu = 0
\label{gc}
\ee
we cast the lagrangian in the
form
$$
\cl_{AH} = - \fr{1}{4} W^{\mu\nu} W_{\mu\nu} + \fr{1}{2} e^2 \rho^2
\ W_\mu \lef [ \fr{-\Box \delta^{\mu\nu} +\del^\mu\del^\nu}{-\Box}
\ri ] W_\nu +\fr{1}{2}\del_\mu \rho \del^\mu \rho - V(\rho)
$$
or
\be
\cl_{AH} =  -\fr{1}{4} W^{\mu\nu}\lef [ 1 + \fr{e^2 \rho^2}{-\Box} \ri ]
W_{\mu\nu}
 + \fr{1}{2} \del_\mu \rho \del^\mu \rho - V(\rho)
\label{lro1}
\ee
In this form, the gauge invariance of the lagrangian becomes explicit.

\spa We can now construct the vortex creation operator appropriate for the
Abelian Higgs model. It is clear from the analysis done in the previous
subsection that a local operator correlation function
can be obtained by adding to $W_{\mu\nu}$
the external field intensity tensor $\bar A_{\mu\nu}$, similarly to
(\ref{2.14}):
$$
<\mu(x)\mu^\dagger(y)>=Z^{-1} \int DW_\mu D\rho \expo \lef [ \fr{1}{4}
(W_{\mu\nu}+ \overbar A_{\mu\nu})\lef [ 1 + \fr{e^2 \rho^2}{-\Box} \ri]
(W^{\mu\nu}+\overbar A^{\mu\nu}) \ri .\ri.
$$
\be
\lef .\lef.
+ \cl_{GF}
+ \fr{1}{2} \del_\mu \rho \del^\mu \rho
 + V(\rho)\ri ]\ri\}
\label{corr}
\ee
Surface independence of the above expression can be shown by applying the
same transformation as in (\ref{2.5}). Also in (\ref{corr}), we can
exchange $\bar A_{\mu\nu}$ for $\tilde B_{\mu\nu}$ using the same
procedure utilized
in the last subsection thereby making surface independence explicit.

\spa From (\ref{corr}), we can infer the form of the quantum vortex operator
in the Abelian Higgs model, namely
\be
\mu (x) = \exp \lef\{ \fr{1}{2} \int d^3z  \tilde B_{\alpha\beta}
 \lef [ 1 + \fr{e^2 \rho^2}{-\Box} \ri]     F^{\alpha\beta}(z)
  \ri \}
\label{muah}
\ee

In what follows, we are going to study the properties of the vortex
operator both in the Maxwell and Abelian Higgs theories.

\subsection{Commutation Rules}

\spa Let us evaluate in this subsection the
commutation rules of $\mu$ which characterize it as
a vortex operator. First of all, consider the commutator with the
topological charge, which in 2+1D happens to be the magnetic flux
along the xy-plane,
$$
Q=\int d^2z J^0 = \int d^2z \epsilon^{ij}\del_i A_j
$$
\be
J^\mu = \epsilon^{\mu\nu\alpha}\del_\nu A_\alpha
\label{2.23}
\ee

\spa Observe that we can write the vortex operator as $\mu \equiv e^A$, where
both in the pure Maxwell theory and the Abelian Higgs model,
we have
\be
 A = - i \pi b \int_{\vec x,L}^\infty d \xi^i
\epsilon_{ij}
 \Pi^j(\xi,x^0)
 \label{expmu}
\ee
where $\Pi^i$ is the momentum canonically conjugated to $A_i$.

\spa Using the fact that $[A,J^0]$ is a c-number, we can write
\be
[J^0(y),\mu(x)] = - \mu(x) [A(x),J^0(y)]
\label{2.24}
\ee
Indeed,
$$
[A(x),J^0(y)] = i 2\pi b \int_{\vec x,L}^\infty d\xi^i
\epsilon_{ij} \epsilon^{kl} \del_k^{(y)}
[\Pi^j (\vec \xi,t),A_l (\vec y,t)]
$$
\be
[A(x),J^0(y)] = - 2\pi b \int_{\vec x,L}^\infty d\xi^i
\del_i^{(\xi)} \delta^2 (\vec \xi - \vec y) = -
2 \pi b \delta^2 (\vec x-\vec y)
\label{2.25}
\ee
where we have used the fact that $[\Pi^j (\vec \xi,t),A_l (\vec y,t)]
= i \delta^j_l \delta^2 (\vec \xi -\vec y)$.
Inserting (\ref{2.25}) in (\ref{2.24}) and integrating over $y$,
we get
\be
[Q,\mu(x)] = 2\pi b \mu(x)
\label{2.26}
\ee
This shows that $\mu$ creates states carrying $2 \pi b$ units of
topological charge (magnetic flux).

\spa There is another commutation relation which characterizes $\mu$ as
a vortex
creation operator. This is the one with the gauge field potential $A_i$.
Once again, following exactly the same procedure above, we obtain
$$
[\mu(\vec x,t), A_i(\vec y,t)] =  2\pi b \mu(\vec x,t)
\int_{\vec x,L}^\infty d\xi^j \epsilon^{ij} \delta^2 (\vec \xi -\vec y)
$$
This can be written as
$$
[\mu(\vec x,t), A_i(\vec y,t)] = \lim_{\rho,\delta \longrightarrow 0}
\oint_{C_x} \epsilon_{ij} d\xi^j arg(\vec \xi -\vec x)
\delta^2(\vec \xi -\vec x)
$$
  By the use of Stokes' theorem, this can be written as
$$
[\mu(x), A_i(y)] =\mu(x) b \int_{T_x(C)} d^2 \vec \xi
\del^{(\xi)}_i [ \delta^2(\vec\xi -\vec y) arg(\vec \xi - \vec x)]
$$
$$ =
\mu(x) b \lef [ \del^{(y)}_i arg(\vec y -\vec x) \Theta(\vec y; T_x(C)) +
 arg(\vec y -\vec x) \int_{T_x(C)} d^2\xi \del^\xi_i \delta^2(\xi - y)
\ri ]
$$
By using Stokes' theorem once more,
the last surface integral above can be transformed in a line integral
which vanishes in the limit when the $\rho$ and $\delta$ cutoffs are
removed
$$
\oint_{C_x} d\xi^j \epsilon^{ij} \delta^2(\xi - y) \tendeo 0
$$
We therefore conclude that the commutator of the $\mu$ operator with
the gauge potential is,
\be
[\mu(x), A_i(y)] =
\mu(x) b  \del^{(y)}_i arg(\vec y -\vec x)
\label{2.27}
\ee
where we have already taken the limit in which the cutoffs $\rho$
and $\delta$ vanish and the Heaviside function in the first term
becomes identically one.
This relation again shows that the  $\mu$ operator does indeed create
a vortex field configuration.

\section{The Quantum Vortex Correlation Functions }

\setcounter{equation}{0}
\subsection{The Maxwell Theory}

\spa Let us evaluate here the euclidean
correlation functions of the vortex operator
introduced in the last section.
Using (\ref{2.21}), we can write
$$
<\mu(x)\mu^\dagger(y)> = Z^{-1} \int DA_\mu \exp \lef \{- \int d^3z
\lef [ \frac{1}{4} F^{\mu\nu}F_{\mu\nu} + \cl_{GF} + \fr{1}{2}
\tilde B_{\mu\nu}(z;x,y)F^{\mu\nu}+ \ri .  \ri .
$$
\be
\lef . \lef .
\fr{1}{4} \B_{\mu\nu}\B^{\mu\nu}
\ri ] \ri \}
\label{3.1}
\ee
where $\tilde B_{\mu\nu}(z;x,y)= i \tilde B_{\mu\nu}(z;x) -
i \tilde B_{\mu\nu}(z;y)$ and $\tilde B_{\mu\nu}(z;x)$
is given by (\ref{2.22}). The $i$ factor in the previous
expression appears after the analytic continuation to the
euclidean space. The last term in the exponent in (\ref{3.1})
is the renormalization factor,
equivalent to the one appearing in (\ref{2.4}),
which will eliminate the unphysical self-interaction of the the line $L$
which occurs in the definition of the $\mu$ operator.

\spa The quadractic functional integral in (\ref{3.1}) can be easily
performed with the help of the euclidean propagator of the
electromagnetic field and the result is
\be
<\mu(x)\mu^\dagger(y)>  = \exp \lef \{\fr{1}{8} \int d^3z d^3z'
\B_{\mu\nu}(z) \B_{\alpha\beta}(z') P^{\mu\nu}_\lambda
P'^{\alpha\beta}_\rho D^{\lambda\rho}(z-z') - \fr{1}{4} \int d^3z
\B^{\mu\nu}\B_{\mu\nu} \ri\}
\label{3.2}
\ee
where $P^{\mu\nu}_\alpha = \del^\mu \delta^\nu_\alpha -
\del^\nu \delta^\mu_\alpha$ and
$$
D^{\lambda\rho} = [ -\Box \delta^{\lambda\rho} + (1-\xi^{-1}) \del^\lambda
\del^\rho ] \lef [ \fr{1}{(-\Box)^2} \ri ]
$$
is the euclidean $A_\mu$ propagator. To comply with the gauge condition
(\ref{gc}) we must actually choose $\xi \rightarrow \infty$.

\spa We immediately see that the second term of $D$ vanishes anyway when
contracted with
the $P'$s and only the gauge invariant first part gives a contribution to
the correlation function. Inserting (\ref{2.22}) in (\ref{3.2}),
contracting the two $P'$s and integrating over $z$ and $z'$ we get
$$
<\mu(x)\mu^\dagger(y)> = \exp \lef \{ \fr {b^2}{2} \sum_{i,j=1}^2 \lambda_i
\lambda_j \lef \{
\int_{x_i}^\infty d\xi_k \epsilon^{k\mu\nu} \int_{x_j}^\infty
d\eta_l \epsilon^{\l\alpha\beta}
 \fr{1}{4}
 \lef [ \del_\mu\del'_\alpha \delta^{\beta\nu} - \del_\mu \del'_\beta
\delta^{\alpha\nu} + \ri .\ri .\ri .
$$
\be
\lef .\lef .\lef .
\del_\nu\del'_\beta \delta^{\alpha\mu} -
\del_\nu\del'_\alpha \delta^{\beta\mu} \ri ] F(\xi -\eta)-
\int_{x_i}^\infty d\xi_i \int_{x_j}^\infty d\eta_i \delta^3(\xi -\eta)
\ri \} \ri \}
\label{3.3}
\ee
where $x_1 \equiv x$, $x_2 \equiv y$, $\lambda_1 \equiv 1$ and
$\lambda_2 \equiv -1$.
The last term is the renormalization factor and
\be
F(x) = \fr{1}{-\Box}_E = \fr{1}{4\pi |x|}
\label{3.4}
\ee
The first four terms above are equal and give, each one, a contribution
\be
\int_{x_i}^\infty d\xi_i \int_{x_j}^\infty d\eta_i \lef [ \delta^{ij}
(-\Box_E) - \del^{(\xi)}_i \del^{(\eta)}_j \ri ] F(\xi -\eta)
\label{eq}
\ee
We see that the first term above, which contains
all the line dependence and therefore the
nonlocality, is canceled by the last one in
(\ref{3.3}). The second one, produces four terms, $F(A-B)$, where $A$ and $B$
correspond to the
limits of the two integrals. Since $\sum_i \lambda_i =0$, however,
only the $F(x_i -x_j)$ contribution to the sum in (\ref{3.3})
survives.
We finally conclude that
\be
<\mu(x)\mu^\dagger(y)> = \exp \lef \{- \fr{b^2}{2} \sum^2_{ij} \lambda_i
\lambda_j F(x_i -x_j) \ri \} =  \exp \lef \{ b^2 F(x-y) -
b^2 F(\epsilon) \ri \}
\label{eq1}
\ee

Introducing the renormalized field
$$
\mu_R = \mu \exp  \{ \fr{b^2}{2} F(\epsilon) \}
$$
we obtain the final result for the vortex correlation function in
Maxwell's theory:
\be
<\mu(x)\mu^\dagger(y)>_R = \exp \lef \{ \fr{b^2}{4\pi |x-y|} \ri \}
\label{3.6}
\ee

 The behavior of this correlation function at infinity shows that there are no
genuine vortex excitations in pure Maxwell theory. This happens because
\be
<\mu\mu^\dagger> \tende |<\mu>|^2 = 1
\label{3.7}
\ee
This relation clearly shows that $<0|\mu> = 1$, indicating that the vortex
states in this case are not orthogonal to the vacuum.

\subsection{The Abelian Higgs Model}

Let us consider now  the nontrivial case of the Abelian Higgs Model,
which is described by the lagrangian given either by (\ref{2.7}) or
(\ref{lro1}).
	Let us concentrate first on the broken phase, where
 $g^2 < 0$,  $<\rho>\equiv
\rho_0 = (4\pi|g|^2/\lambda) $ and the gauge field acquires a mass
 $M= e \rho_0$ after the shift $\rho \rightarrow \rho +\rho_0$ is
 performed.
The vortex correlation function is now given by
$$
<\mu(x)\mu^\dagger(y)>=Z^{-1} \int DW_\mu D\rho \expo \lef [ \fr{1}{4}
(W_{\mu\nu}+ \tilde B_{\mu\nu})
\lef [ 1 + \fr{e^2 (\rho + \rho_0)^2}{-\Box_E} \ri]  \ri . \ri .
$$
\be
\lef .\lef .
(W^{\mu\nu}+\tilde B^{\mu\nu}) + \cl_{GF}
+ \fr{1}{2} \del_\mu \rho \del^\mu \rho
 + V(\rho)\ri ]\ri\}
 = \exp \{ \Lambda(x,y) - S_R [\tilde B_{\mu\nu} ] \}
\label{corr1}
\ee

In this expression, $\Lambda(x,y)$ is the sum of all Feynman graphs with
the field $\tilde B_{\mu\nu}$ in the external legs and $S_R [\tilde B_
{\mu\nu} ]$ is the self-coupling of this external field.
 We are going to evaluate (\ref{corr1}) by means of a loop expansion.
 Furthermore, we are
going to consider the long distance limit of (\ref{corr1}). In this
limit, it can be shown \cite{vorcorr,an} that only
two legs graphs contribute to  $\Lambda(x,y)$.

The relevant vertex for the leading term (0-loop)
in this expansion comes from
the linear coupling of
  $\tilde B_{\mu\nu}$
with the gauge
field $W_\mu$,
 which can be inferred from expression (\ref{corr1}), namely
\be
\fr{1}{2} \tilde B_{\mu\nu} P^{\mu\nu}_\alpha \lef [1+ \fr{M^2}{-\Box_E}\ri ]
W^\alpha
\label{vert}
\ee

 Under the above mentioned conditions, the leading contribution to
 $\Lambda(x,y)$ is given by the graph of Fig. 3 and the correlation function
is given by
$$
<\mu(x)\mu^\dagger(y)> = \exp \lef \{ \fr{1}{8}\int d^3z d^3z'
\tilde B^{\mu\nu}(z)
\tilde B^{\alpha\beta}(z')  P^{\mu\nu}_\lambda P'^{\alpha\beta}
_\rho \lef [1 + \fr{M^2}{-\Box_E}\ri] \lef [1 + \fr{M^2}{-\Box_E}\ri]\ri .
$$
\be
\lef .
D_M^{\lambda\rho} (z-z')- \fr{1}{4} \int d^3z \B_{\mu\nu}
\lef [1 + \fr{M^2}{-\Box_E}\ri]
\B^{\mu\nu}
\ri \}
\label{3.9}
\ee

 In this expression, $D_M^{\lambda\rho} (z-z')$ is the massive gauge field
propagator, which is given by
\be
 D_M^{\alpha\beta}= \fr{\delta^{\alpha\beta}}{-\Box_E + M^2} + gauge\  terms
\label{3.10}
\ee
and the last term is the self-interaction of the external field.

We may evaluate (\ref{3.9}) by following exactly the same steps as we did
in the last subsection, Eqs.(\ref{3.2}) to (\ref{eq}). The only difference
is that now $F(x)$ in (\ref{3.3}), (\ref{3.4}) and (\ref{eq}) is exchanged by
$$
F_M(x) = \lef [1 + \fr{M^2}{-\Box_E}\ri] \lef [ \fr{1}{-\Box_E + M^2}\ri ]
\lef [1 + \fr{M^2}{-\Box_E}\ri]
$$
\be
= \fr{1}{-\Box_E} + \fr{M^2}{(-\Box_E)^2} = \lim_{m \rightarrow 0}
\fr{1}{4\pi |x|} + \fr{M^2}{4\pi} \lef [\fr{1}{m} -\fr{|x|}{2}\ri ]
\label{fm}
\ee
and the self-interaction term, the last one in (\ref{3.9}), contains the
$M^2$ dependent piece.
As a consequence, the last
term in (\ref{3.3}), which contains a delta function
becomes multiplied by $1+M^2/-\Box_E$.

\spa In (\ref{fm}), $m$ is a mass regulator introduced in order to give
meaning to the inverse Fourier transform of $\fr{1}{k^4}$, which appears
corresponding to $\fr{1}{(-\Box_E)^2}$.

\spa Observe that in the same way as in the case of pure
Maxwell theory, the first term in
(\ref{eq}) which contains all the nonlocal dependence
is exactly canceled by the renormalization term.
Following the same steps as we did from (\ref{eq1}) to (\ref{3.6})
it is easy to see that all the $m$ dependence vanishes because
$\sum_i \lambda_i =0$ and we get
\be
<\mu(x)\mu^\dagger(y)>_R \tende \exp \lef \{ - {\cal M}|x-y| +
\fr{b^2}{4\pi |x-y|} \ri \}
\label{corr2}
\ee
In this expression, ${\cal M}=\fr{b^2 M^2}{8\pi} = \fr{b^2 e^2 \rho_0^2}{8\pi}$
is the quantum vortex mass in
the order of approximation we are working at. We now see that the vortex
operator indeed creates genuine excitations which are orthogonal to
the vacuum. This follows from the fact that $<\mu\mu^\dagger> \tende 0$
and hence $<\mu> = <0|\mu> = 0$.

\spa In the symmetric phase, where $\rho_0 \rightarrow 0$ the vortex mass
${\cal M}$ vanishes and only the second term contributes to (\ref{corr2}).
The correlation function has the same long distance behavior as the exact
one in pure Maxwell theory. This indicates that, as expected, there are
no genuine vortex excitations in the symmetric phase.

\vfill
\eject

\section{Concluding Remarks}

\spa There are many advantages in the new formulation for the quantization
of three-dimensional vortices presented here. The vortex operator is
explicitly surface invariant. There are no cutoffs appearing in the
definition of the operator. Even in the case of the Abelian Higgs model there
are only gauge invariant degrees of freedom appearing in the definition
of the operator, in such a way that gauge invariance is always explicit.
Perhaps the nicest feature of this formulation, however, is the fact that
a mass expansion is not required for the obtainment of the vortex correlation
function in the nontrivial case of the Abelian Higgs model. The mass expansion
was a common characteristic of the previous formulation. Also there is an
exact cancelation of all the nonlocal terms already in lowest order in the
loop expansion.

\spa There are many possible applications for the formulation developed here.
We can envisage the case of vortices in superconductors and maybe also
superfluids. It would be interesting anyway to study the temperature
dependence of the correlation functions studied here in connection with these
applications.

\spa We have also established a generalization of the present formulation
for the case of quantum cosmic strings in 3+1D which is going to be published
elsewhere. In any of the above mentioned cases, however, in order to obtain
quantities like the scattering amplitudes
of topological excitations as well as
cross sections for processes involving these excitations, the knowledge of
an asymptotic theory for the corresponding creation operators will be needed.
This is so far not known and is a very interesting field of research.

\vfill
\eject

\vfill
\eject

\centerline{\Large\bf Figure Captions}

\bigskip

Fig. 1 - Surface used in the definition
of the external field $A_{\mu\nu}(z;x)$ and the operator $\mu$

Fig. 2 - The surface $\tilde T_x(C)$

Fig. 3 - Leading graph contributing to the vortex correlation function.
The external legs are $\tilde B_{\mu\nu}$'s. The wavy line is a $W_\mu$
propagator. The vertex is defined by \ref{vert}.


\begin{thebibliography}{99}



\bibitem{vor} E.C.Marino, {\it Phys. Rev.} {\bf D38} (1988) 3194

\bibitem{rede}J.Fr\" ohlich and P.A.Marchetti, {\it Lett. Math. Phys.}
{\bf 16} (1988) 347; {\it Commun. Math. Phys. } {\bf 121} (1989) 177;
J.Fr\" ohlich, F.Gabiani and P.A.Marchetti, in ``Proceedings, Banff
Summer School in Theoretical Physics'' (H.C.Lee, Ed.)

\bibitem{cont} G.W.Semenoff and P.Sodano, {\it Nucl. Phys.} {\bf B328}
(1989) 753; M.L\"uscher, {\it Nucl. Phys.} {\bf B326} (1989) 557;
R.Jackiw and S.Y.Pi, {\it Phys. Rev.} {\bf D42} (1990) 3500;
A.Kovner, B.Rosenstein and D.Eliezer, {\it Nucl. Phys.} {\bf B350}
(1991) 325; {\it Mod. Phys. Lett.} {\bf A5} (1990) 2733;
V.F.M\" uller, {\it Z.Phys.} {\bf C51} (1991) 665

\bibitem{vorcorr} E.C.Marino, G.C.Marques, R.O.Ramos and
J.E.Stephany Ruiz, {\it Phys. Rev.} {\bf D45} (1992) 3690

\bibitem{an}  E.C.Marino, {\it Ann. of Phys. (NY) } {\bf 224 } (1993) 225

\bibitem{man} S.Mandelstam  {\it Phys. Rev.} {\bf D11} (1975) 3026

\bibitem{evora} E.C.Marino, {\it ``Dual Quantization of Solitons''} in
Proceedings of the NATO Advanced Study Institute, Applications of Statistical
and Field Theory Methods in Condensed Matter, (D.Baeriswyl, A.Bishop and
J.Carmelo, Eds.) Plenum, NY, 1992.

\end{thebibliography}
\end{document}